\documentclass[conference,a4paper]{IEEEtran}

\usepackage{cite}

\usepackage{graphicx,color,epsfig,rotating,subfigure}
\usepackage{amsfonts,amsmath,amssymb}
\usepackage{algorithm,algorithmic}
\usepackage{subfigure}

\long\def\comment#1{}

\newfont{\bbb}{msbm10 scaled 700}

\newfont{\bb}{msbm10 scaled 1100}
\newcommand{\CC}{\mbox{\bb C}}
\newcommand{\PP}{\mbox{\bb P}}

\newcommand{\ZZ}{\mbox{\bb Z}}
\newcommand{\FF}{\mbox{\bb F}}

\newcommand{\av}{{\bf a}}
\newcommand{\bv}{{\bf b}}
\newcommand{\cv}{{\bf c}}
\newcommand{\dv}{{\bf d}}

\newcommand{\rv}{{\bf r}}
\newcommand{\sv}{{\bf s}}
\newcommand{\tv}{{\bf t}}

\newcommand{\wv}{{\bf w}}
\newcommand{\vv}{{\bf v}}
\newcommand{\xv}{{\bf x}}
\newcommand{\yv}{{\bf y}}
\newcommand{\zv}{{\bf z}}
\newcommand{\zerov}{{\bf 0}}
\newcommand{\onev}{{\bf 1}}

\newcommand{\Am}{{\bf A}}
\newcommand{\Bm}{{\bf B}}
\newcommand{\Cm}{{\bf C}}
\newcommand{\Dm}{{\bf D}}

\newcommand{\Fm}{{\bf F}}
\newcommand{\Gm}{{\bf G}}
\newcommand{\Hm}{{\bf H}}
\newcommand{\Id}{{\bf I}}

\newcommand{\Lm}{{\bf L}}
\newcommand{\Mm}{{\bf M}}

\newcommand{\Qm}{{\bf Q}}

\newcommand{\Sm}{{\bf S}}
\newcommand{\Tm}{{\bf T}}
\newcommand{\Um}{{\bf U}}
\newcommand{\Wm}{{\bf W}}
\newcommand{\Vm}{{\bf V}}
\newcommand{\Xm}{{\bf X}}
\newcommand{\Ym}{{\bf Y}}
\newcommand{\Zm}{{\bf Z}}

\newcommand{\Cc}{{\cal C}}

\newcommand{\Lc}{{\cal L}}

\newcommand{\Nc}{{\cal N}}

\newcommand{\Vc}{{\cal V}}

\newcommand{\alphav}{\hbox{\boldmath$\alpha$}}

\newcommand{\lambdav}{\hbox{\boldmath$\lambda$}}

\newcommand{\trace}{{\hbox{tr}}}

\newcommand{\SNR}{{\sf SNR}}

\newcommand{\herm}{{\sf H}}

\newtheorem{theorem}{Theorem}
\newtheorem{lemma}{Lemma}
\newtheorem{corollary}{Corollary}

\newcommand{\argmin}{\operatornamewithlimits{argmin}}

\begin{document}

\sloppy

\title{Structured Lattice Codes for $2 \times 2 \times 2$ MIMO Interference Channel}

\author{
  \IEEEauthorblockN{Song-Nam~Hong}
  \IEEEauthorblockA{Dep. of Electrical Engineering\\
    University of Southern California\\
    CA, USA\\
    Email: songnamh@usc.edu}
  \and
  \IEEEauthorblockN{Giuseppe~Caire}
  \IEEEauthorblockA{Dep. of Electrical Engineering\\
    University of Southern California\\
    CA, USA\\
    Email: caire@usc.edu}
}



\maketitle

\begin{abstract}
We consider the $2 \times 2 \times 2$ multiple-input multiple-output interference channel
where two source-destination pairs wish to communicate with the aid of two intermediate relays.
In this paper, we propose a novel lattice strategy called Aligned Precoded Compute-and-Forward (PCoF).
This scheme consists of two phases:
1) Using the CoF framework based on signal alignment we transform the Gaussian network into a {\em deterministic} finite field network.
2) Using linear precoding (over finite field) we eliminate the end-to-end interference in the finite field domain.
Further, we exploit the algebraic structure of lattices to enhance the performance at finite SNR,
such that beyond a degree of freedom result (also achievable by other means). We can also
show that Aligned PCoF outperforms time-sharing in a range of reasonably moderate SNR, with increasing gain as SNR increases.
\end{abstract}

\section{Introduction}\label{sec:int}

In recent years, significant progress has been made on the understanding of the theoretical limits of wireless communication networks.
In \cite{Avestimehr}, the capacity of multiple multicast network (where every destination desires all messages)
is approximated within a constant gap independent of SNR and of the realization of the channel coefficients.
Also, for multiple flows over a single hop, new capacity approximations were obtained
in the form of degrees of freedom (DoF), generalized degrees of freedom (GDoF), and $O(1)$ approximations \cite{Cadambe08,Gou09,Jafar10}.
Yet, the study of multiple flows over multiple hops remains largely unsolved.
The $2 \times 2 \times 2$ Gaussian interference channel (IC) has received much attention recently, being one of the fundamental building blocks to
characterize the DoFs of two-flows networks\cite{Shomorony}
One natural approach is to consider this model as a cascade of two ICs.
In \cite{Simeone}, the authors apply the Han-Kobayashi scheme \cite{Han} for the first hop to split each message into private and common parts.
Relays can cooperate using the shared information (i.e., common messages) for the second hop, in order to improve the data rates.
This approach is known to be highly suboptimal at high signal-to-noise ratios (SNRs), since two-user IC can only achieve 1 DoF.
In \cite{Cadambe}, Cadembe and Jafar show that $\frac{4}{3}$ DoF is achievable by viewing each hop as an X-channel.
This is accomplished using the {\em interference alignment} scheme for each hop.
Recently, the optimal DoF was obtained in \cite{Gou} using {\em aligned interference neutralization},
which appropriately combines interference alignment and interference neutralization. Also, there was the recent extension to the $K \times K \times K$ Gaussian IC in \cite{Shomorony1}, achieving the optimal $K$ DoF using {\em aligned network diagonalization}.


\begin{figure}
\centerline{\includegraphics[width=8cm]{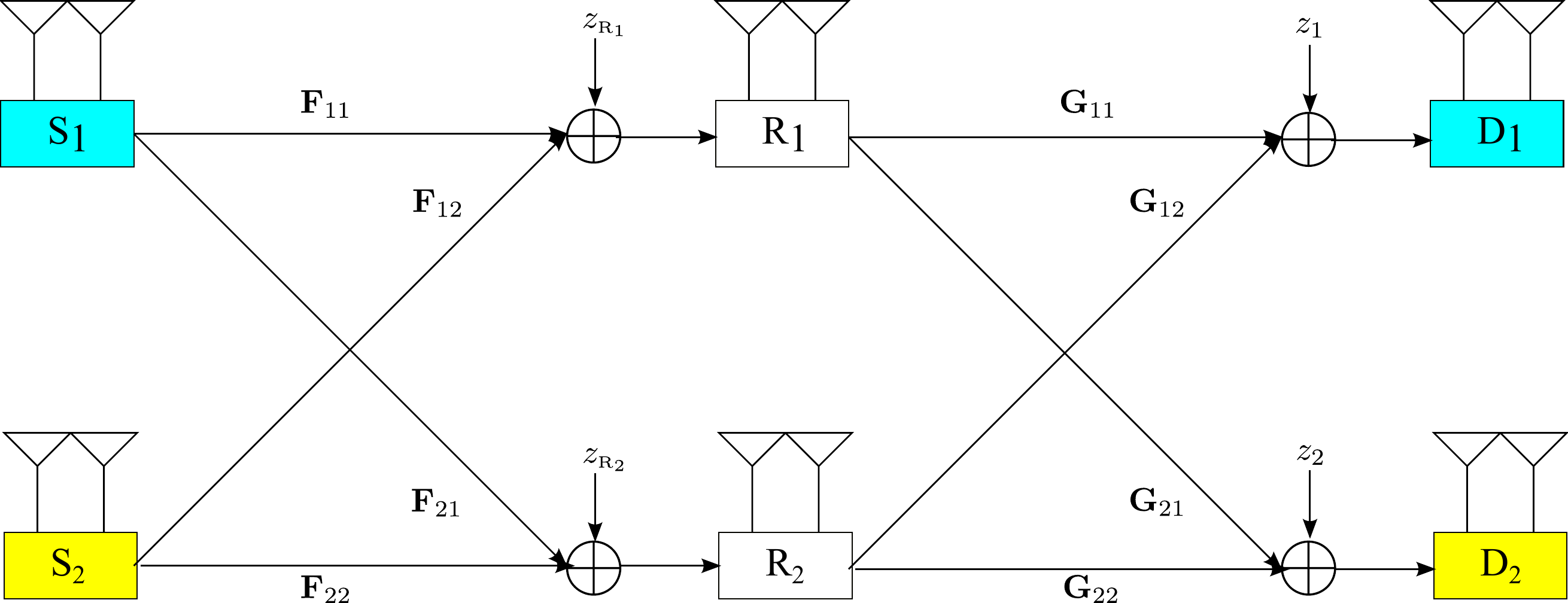}}
\caption{$2\times 2\times 2$ MIMO Gaussian interference channel.}
\label{222model}
\end{figure}


We consider the $2 \times 2 \times 2$ multiple-input multiple-output (MIMO) IC as shown in Fig.~\ref{222model} consisting of two sources,
two relays, and two destinations.
All nodes have $M$ multiple antennas.
Each source $k$ has a message for its intended destination $k$, for $k=1,2$.
In the first hop, a block of $n$ channel uses of the discrete-time complex MIMO IC is described by
\begin{equation}
\left[
  \begin{array}{c}
    \underline{\Ym}_{\mbox{\tiny{R}}_{1}} \\
    \underline{\Ym}_{\mbox{\tiny{R}}_{2}} \\
  \end{array}
\right]=\left[
                                             \begin{array}{cc}
                                               \Fm_{11} & \Fm_{12} \\
                                               \Fm_{21} & \Fm_{22} \\
                                             \end{array}
                                           \right]\left[
  \begin{array}{c}
    \underline{\Xm}_{1} \\
    \underline{\Xm}_{2} \\
  \end{array}
\right]+\left[
  \begin{array}{c}
    \underline{\Zm}_{\mbox{\tiny{R}}_{1}} \\
    \underline{\Zm}_{\mbox{\tiny{R}}_{2}} \\
  \end{array}
\right]\label{model:1hop}
\end{equation}where the matrices $\underline{\Xm}_{k}$~\footnote{ We use ``underline" to denote the matrices whose horizontal dimension (column index) denotes ``time" and vertical dimension (row index) runs across the antennas. For any matrix $\underline{\Xm}$, we denote $\underline{\xv}_{\ell}$ by the $\ell$-th row of $\underline{\Xm}$.}
and $\underline{\Ym}_{\mbox{\tiny{R}}_{k}}$ contain, arranged by rows, the $k$-th source channel input sequences
$\underline{\xv}_{k,\ell} \in \CC^{1 \times n}$ for $\ell = 1,\ldots, M$,
the $k$-th relay channel output sequences $\underline{\yv}_{\mbox{\tiny{R}}_{k,\ell}} \in \CC^{1 \times n}$,
and where $\Fm_{j k} \in \CC^{M \times M}$ denotes the channel matrix from source $k$ to $j$.
In the second hop, a block of $n$ channel uses of the discrete-time complex MIMO IC is described by
\begin{equation}
\left[
  \begin{array}{c}
    \underline{\Ym}_{1} \\
    \underline{\Ym}_{2} \\
  \end{array}
\right]=\left[
                                             \begin{array}{cc}
                                               \Gm_{11} & \Gm_{12} \\
                                               \Gm_{21} & \Gm_{22} \\
                                             \end{array}
                                           \right]\left[
  \begin{array}{c}
    \underline{\Xm}_{\mbox{\tiny{R}}_{1}} \\
    \underline{\Xm}_{\mbox{\tiny{R}}_{2}} \\
  \end{array}
\right]+\left[
  \begin{array}{c}
    \underline{\Zm}_{1} \\
    \underline{\Zm}_{2} \\
  \end{array}
\right]\label{model:2hop}
\end{equation}
where  the matrices $\underline{\Xm}_{\mbox{\tiny{R}}_{k}}$ and $\underline{\Ym}_{k}$ contain the $k$-th
relay channel input sequences $\underline{\xv}_{\mbox{\tiny{R}}_{k,\ell}} \in \CC^{1 \times n}$,
the $k$-th destination channel output sequences  $\underline{\yv}_{k,\ell} \in \CC^{1 \times n}$,
and where $\Gm_{k j} \in \CC^{M \times M}$ denotes the channel matrix between relay $j$ and destination $k$.
The matrix $\underline{\Zm}_{k}$ (or $\underline{\Zm}_{\mbox{\tiny{R}}_{k}}$) contains i.i.d. Gaussian noise samples $\sim \Cc\Nc(0,1)$. We assume that the elements of
$\Fm_{j k}$ and $\Gm_{kj}$ are drawn i.i.d. according to a continuous distribution (i.e., Gaussian distribution).
The channel matrices are assumed to be constant over the whole block of length $n$ and known to all nodes.
Also, we consider a sum-power constraint equal to $M\SNR$ at each transmitter.

In this paper, we propose a novel lattice strategy named Aligned Precoded Compute-and-Forward (PCoF). CoF makes use of lattice codes so that each receiver can reliably decode a linear combination with integer coefficients of the interfering codewords. Thanks to the fact that lattices are modules over the ring of integers, the linear combinations translates directly into a linear combination of messages over a suitable finite field. In this way, each hop is transformed into a {\em deterministic} noiseless finite field IC. The end-to-end interferences in the finite field domain are eliminated by distributed precoding (over finite field) at relays. Using this framework, we characterize a {\em symmetric} sum rate
and prove that $2M-1$ DoF is achievable by lattice coding (this DoF result is proven in
\cite{Gou} without resorting to CoF and lattice coding). Further, we use the lattice codes algebraic structure in  order to obtain also good performance
at finite SNRs. We use {\em integer-forcing receiver} (IFR) of \cite{Zhan} in order to minimize the impact of noise boosting at the receivers,
and {\em integer-forcing beamforming} (IFB), proposed by the authors in \cite{Hong}, in order to minimize the power penalty at the transmitters.
We provide numerical results showing that Aligned PCoF outperforms
time-sharing even at reasonably moderate SNR, with increasing performance gain as SNR increases.

\section{Preliminaries}

In this section we provide some basic definitions and results that will be extensively used in the sequel.
%

\subsection{Nested Lattice Codes}\label{subsec:NLC}

Let $\ZZ[j]$ be the ring of Gaussian integers and $p$ be a prime. Let $\oplus$ denote the addition over $\FF_{q}$ with $q=p^2$, and let $g: \FF_{q} \rightarrow \CC$ be the natural mapping of $\FF_{q}$ onto $\{a+jb: a,b \in \ZZ_{p}\} \subset \CC$. We recall the nested lattice code construction given in \cite{Nazer}.
Let $\Lambda = \{ \underline{\lambdav} = \underline{\zv} \Tm : \underline{\zv} \in \ZZ^n[j]\}$ be a lattice in $\CC^n$,
with full-rank generator matrix $\Tm \in \CC^{n \times n}$.
Let $\Cc  = \{ \underline{\cv} = \underline{\wv} \Gm : \wv \in \FF_{q}^r \}$ denote a linear code over $\FF_{q}$ with block length $n$ and dimension $r$, with generator matrix $\Gm$. The lattice $\Lambda_1$ is defined through ``construction A'' (see \cite{Erez2004} and references therein) as
\begin{equation} \label{construction-A}
\Lambda_1 = p^{-1} g(\Cc) \Tm + \Lambda,
\end{equation}
where $g(\Cc)$ is the image of $\Cc$ under the mapping $g$ (applied component-wise). It follows that $\Lambda \subseteq \Lambda_1 \subseteq p^{-1} \Lambda$ is a chain of nested lattices, such that
$|\Lambda_1/\Lambda| = p^{2r}$ and $|p^{-1} \Lambda/\Lambda_1| = p^{2(n - r)}$.

For a lattice $\Lambda$ and $\underline{\rv} \in \CC^n$,
we define the lattice quantizer $Q_{\Lambda}(\underline{\rv}) = \argmin_{\underline{\lambdav} \in \Lambda}\|\underline{\rv} - \underline{\lambdav} \|^2$, the Voronoi region $\Vc_\Lambda = \{\underline{\rv} \in \CC^{n}: Q_{\Lambda}(\underline{\rv}) = \underline{\zerov}\}$
and $[\underline{\rv}] \mod \Lambda = \underline{\rv} - Q_{\Lambda}(\underline{\rv})$.
For $\Lambda$ and $\Lambda_1$ given above, we define the lattice code
$\Lc = \Lambda_{1} \cap \Vc_\Lambda$ with rate $R = \frac{1}{n} \log |\Lc| = \frac{r}{n}\log{q}$. Construction A provides a {\em natural labeling}
of the codewords of $\Lc$ by the information messages $\underline{\wv} \in \FF^r_{q}$.  Notice that the set $p^{-1} g(\Cc)\Tm$ is a {\em system of coset representatives}
of the cosets  of $\Lambda$ in $\Lambda_1$. Hence, the natural labeling function  $f : \FF^r_{q} \rightarrow \Lc$ is defined by $f(\underline{\wv}) = p^{-1} g(\underline{\wv} \Gm)\Tm \mod \Lambda$.

\subsection{Compute-and-Forward}\label{subsec:CoF}

We recall here the CoF scheme of \cite{Nazer}.
Consider a  $2$-user Gaussian multiple access channel (MAC) with $M$ antennas at each transmitter and at the receiver, represented by
\begin{equation}
\underline{\Ym} = \Hm\Cm\underline{\Xm} + \underline{\Zm}
\end{equation} where $\Hm \in \CC^{M \times M}$, $\Cm \in \ZZ[j]^{M \times 2M}$, and $\underline{\Zm}$ contains i.i.d. Gaussian noise samples $\sim \Cc\Nc(0,1)$. This particular form of channel matrix $\Hm\Cm$ will be widely considered in this paper, as a consequence of signal alignment. All users make use of the same nested lattice codebook $\Lc=\Lambda_{1} \cap \Vc_{\Lambda}$, where $\Lambda$ has second moment $\sigma_{\Lambda}^{2} \triangleq \frac{1}{n\mbox{Vol}(\Vc_\lambda)}\int_{\Vc_\lambda}\|\underline{\rv}\|^2 d\underline{\rv}=\SNR_{\mbox{\tiny{eff}}}$.
Each user $k$ encodes $M$ information messages $\underline{\wv}_{k,\ell} \in \FF_{q}^{r}$ into the corresponding codeword $\underline{\tv}_{k,\ell}=f(\underline{\wv}_{k,\ell})$
and produces its channel input according to
\begin{equation}
\underline{\xv}_{k,\ell}= [\underline{\tv}_{k,\ell} + \underline{\dv}_{k,\ell}] \mod \Lambda,
\end{equation}
for $\ell = 1,\ldots, M$,
where the {\em dithering sequences} $\underline{\dv}_{k,\ell}$'s mutually independent across the users and the messages,
uniformly distributed over $\Vc_{\Lambda}$, and known to the receiver.
Notice that $\underline{\xv}_{k,\ell}$ is the $\ell$-th row of $\underline{\Xm}_k$, $k = 1,2$.
The decoder's goal is to recover an integer linear combination $\underline{\sv} = [\bv^{\herm}\Cm\underline{\Tm}] \mod \Lambda$, with some integer vector $\bv \in \ZZ[j]^{M \times 1}$. Since $\Lambda_{1}$ is a $\ZZ[j]$-module (closed under liner combinations with Gaussian integer coefficients), then $\underline{\sv} \in \Lc$.
Letting $\hat{\underline{\sv}}$ be decoded codeword, we say that a computation rate $R$ is achievable for this setting if there exists sequences of lattice codes $\Lc$
of rate $R$ and increasing block length $n$, such that the decoding error probability
satisfies $\lim_{n \rightarrow \infty}\PP(\hat{\underline{\sv}}\neq \underline{\sv})=0$.

In the scheme of \cite{Nazer}, the receiver computes
\begin{eqnarray}
\hat{\underline{\yv}} &=& \left[\alphav^{\herm} \underline{\Ym} - \bv^{\herm}\Cm\underline{\Dm}\right] \mod \Lambda\nonumber\\
&=& \left[\underline{\sv} + \underline{\zv}_{\mbox{\tiny{eff}}}(\Hm\Cm,\bv,\alphav)\right] \mod \Lambda\label{eq:cof}
\end{eqnarray}where $\alphav \in \CC^{M\times 1}$ and
$\underline{\zv}_{\mbox{\tiny{eff}}}(\Hm\Cm,\bv,\alphav) = (\alphav^{\herm}\Hm-\bv^{\herm})\Cm\underline{\Um}+\alphav^{\herm}\underline{\Zm}$ denotes the {\em effective noise}, including the non-integer self-interference (due to the fact that $\alphav^{\herm}\Hm \notin \ZZ[j]^{1 \times M}$ in general) and the additive Gaussian noise term. The scaling, dither and modulo-$\Lambda$ operation in (\ref{eq:cof}) is referred to as the {\em CoF receiver mapping}.
Choosing $\alphav^{\herm}=\bv^{\herm}\Hm^{-1}$, the variance of the effective noise is given by $\sigma^{2}_{\mbox{\tiny{exact}}} = \|(\Hm^{-1})^{\herm}\bv\|^2$.
This scheme is known as {\em exact} IFR \cite{Zhan}. In this way, the non-integer penalty of CoF is completely eliminated.
More in general, the performance can be improved especially at low SNR
by  minimizing the variance of $\underline{\zv}_{\mbox{\tiny{eff}}}(\Hm,\bv,\alphav)$ with respect to $\alphav$. In this case, we obtain:
\begin{eqnarray}
\sigma^{2}(\Hm\Cm,\bv)
&=& \bv^{\herm}\Cm(\SNR_{\mbox{\tiny{eff}}}^{-1}\Id+\Cm^{\herm}\Hm^{\herm}\Hm\Cm)^{-1}\Cm^{\herm}\bv.
\end{eqnarray} Since $\alphav$ is uniquely determined by $\Hm\Cm$ and $\bv$, it will be omitted in the following, for the sake of notation simplicity. From \cite{Nazer}, we know that by applying lattice decoding to $\hat{\underline{\yv}}$ given in (\ref{eq:cof}) the following computation rate is achievable:
\begin{equation}
R(\Hm,\bv,\SNR_{\mbox{\tiny{eff}}}) = \log^{+}(\SNR_{\mbox{\tiny{eff}}}/\sigma^{2}(\Hm\Cm,\bv)) \label{eq:cofrate} 
\end{equation}where $\log^{+}(x)\triangleq \max\{\log(x),0\}$. Also, the receiver can reliably
decode $M$ linear combinations $\underline{\Sm}=[\Bm\Cm\underline{\Tm}] \mod \Lambda$ with integer coefficient vectors
$\{\bv_{\ell}^{\herm}:\ell=1,\ldots,M\}$ (i.e., the $\ell$-th row of $\Bm$) if
\begin{equation}
R\leq \min_{\ell}\{R(\Hm,\bv_{\ell},\SNR_{\mbox{\tiny{eff}}})\}\triangleq R(\Hm,\Bm,\SNR_{\mbox{\tiny{eff}}}). \label{cofrate}
\end{equation} Using the lattice encoding linearity, the corresponding $M$ linear combinations over $\FF_{q}$ for the messages
are obtained as
\begin{eqnarray}
\underline{\Um} &=& g^{-1}([\Bm] \mod p\ZZ[j])g^{-1}([\Cm] \mod p\ZZ[j])\underline{\Wm}\nonumber\\
&\stackrel{(a)}{=}& [\Bm]_{q}[\Cm]_{q}\underline{\Wm}, \label{cofeq}
\end{eqnarray} where we
use the notation $[\Bm]_{q}  \triangleq g^{-1}([\Bm] \mod p\ZZ[j])$.

\begin{figure}
\centerline{\includegraphics[width=8cm]{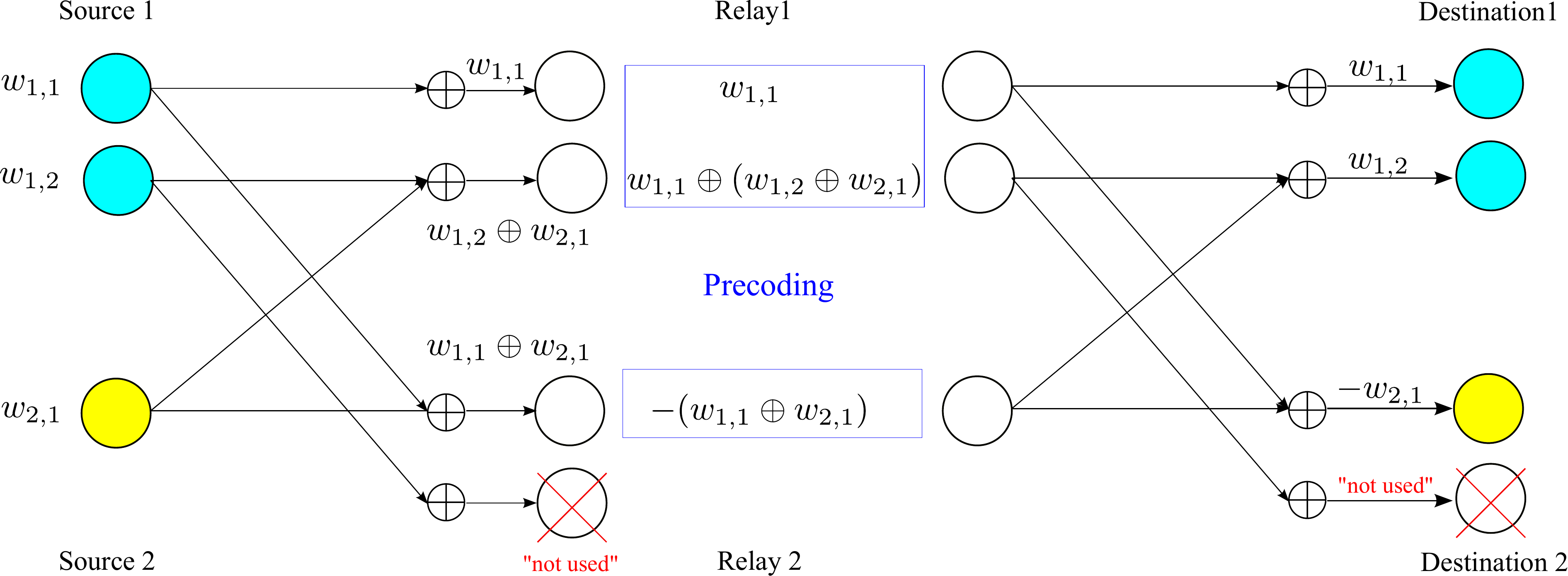}}
\caption{A {\em deterministic} noise-free $2\times 2 \times 2$ finite field interference channel.}
\label{field}
\end{figure}
\section{Aligned PCoF}
In this section, we propose a novel lattice strategy called ``Aligned" Precoded CoF (PCoF). This consists of two phases: 1) The CoF framework transforms a Gaussian network into a finite field network. 2) A linear precoding scheme is used over finite field to eliminate the end-to-end interferences (see Fig.~\ref{field}).
While using the CoF framework, the main performance bottleneck consists of the non-integer penalty, which ultimately limits the performance of CoF scheme
in the high SNR regime \cite{Niesen1}.  To overcome this bottleneck, we employ {\em signal alignment}
in order to create an ``aligned" channel matrix for which exact integer forcing is possible, as seen in Section \ref{subsec:CoF}. Namely, we use
alignment precoding matrices $\Vm_1$ and $\Vm_2$ at the two sources such that
\begin{equation}
\left[
  \begin{array}{cc}
    \Fm_{k1}\Vm_{1} & \Fm_{k2}\Vm_{2} \\
  \end{array}
\right] = \Hm_{\mbox{\tiny{R}}_{k}}\Cm_{\mbox{\tiny{R}}_{k}},
\end{equation}
where $\Hm_{\mbox{\tiny{R}}_{k}} \in \CC^{M\times M}$ and $\Cm_{\mbox{\tiny{R}}_{k}} \in \ZZ[j]^{M\times 2M}$.
However, the precoding over $\CC$ may produce a power penalty due to the non-unitary nature of the alignment matrices, and this  can degrade
the performance at finite SNR. In order to counter this effect, we use the concept of IFB (see \cite{Hong}).
The main idea is that $\Vm_{k}$ can be pre-multiplied (from the oft) by some appropriately chosen
full-rank integer matrix $\Am_{k}$ since its effect can be undone by precoding over $\FF_{q}$, using $[\Am_{k}]_{q}$.
Then, we can optimize the integer matrix in order to minimize the power penalty.
The detailed procedures of Aligned PCoF are given in the following sections.

\subsection{CoF framework based on signal alignment}

In this section we show how to turn any 2-user MIMO IC into a noiseless finite field IC using the CoF framework.
We focus on the first hop of our $2 \times 2 \times 2$ network since the same scheme is straightforwardly used for the second hop.
Consider the MIMO IC in (\ref{model:1hop}).
Let $\{\underline{\wv}_{1,\ell} \in \FF_{q}^{r}: \ell=1,\ldots,M\}$ denote the messages of  source 1 and $\{\underline{\wv}_{2,\ell} \in \FF_{q}^{r}: \ell=1,\ldots,M-1\}$ denote the messages of source 2. All transmitters make use of the same nested lattice codebook $\Lc=\Lambda_{1} \cap \Vc_\Lambda$, where $\Lambda$ has the second moment $\sigma_{\Lambda}^{2}=\SNR_{\mbox{\tiny{eff}}}$. Also, we let
$\Vm_{1}=[\vv_{1,1} \cdots \vv_{1,M}] \in \CC^{M \times M}$ and $\Vm_{2}=[\vv_{2,1} \cdots \vv_{2,M-1}]  \in \CC^{M \times M-1}$ denote the precoding matrices used at sources 1 and 2, respectively. They are chosen to satisfy the {\em alignment conditions}, given by
\begin{eqnarray}
\Fm_{11}\vv_{1,k+1} &=& \Fm_{12}\vv_{2,k}\\
\Fm_{21}\vv_{1,k} &=& \Fm_{22}\vv_{2,k}\label{cond:ALI}
\end{eqnarray} for $k=1,\ldots,M-1$. The feasibility of the above conditions was shown in \cite{Gou} for any $M$.

Let $\Am_{1} \in \ZZ[j]^{M \times M}$ and $\Am_{2} \in \ZZ[j]^{M-1 \times M-1}$ denote the full rank integer matrices. They will be optimized to minimize the power penalty in Section \ref{sec:finite}. Each source $k$ precodes its messages over $\FF_{q}$ as
\begin{equation}
\underline{\Wm}'_{k} = [\Am_{k}]_{q}^{-1}\underline{\Wm}_{k}, \;\;\;\; k=1,2. \label{eq:precoding}
\end{equation}
Then, the precoded messages are encoded using the nested lattice codes. Finally, the channel input sequences are given by the rows of:
\begin{equation}
\underline{\Xm}_{k} = \Vm_{k}\Am_{k}\underline{\Tm}_{k} = \Vm_{k}\underline{\Tm}'_{k},
\end{equation}
where $\underline{\tv}_{k,\ell} = [f(\underline{\wv}'_{k,\ell})+\underline{\dv}_{k,\ell}] \mod \Lambda$, and where
$\underline{\Tm}'_{k} = \Am_{k}\underline{\Tm}_{k}$.
Due to the power constraint, each source $k$ must satisfy
\begin{equation}
\SNR_{\mbox{\tiny{eff}}}\trace(\Vm_{k}\Am_{k}\Am_{k}^{\herm}\Vm_{k}^{\herm}) \leq M\SNR.
\end{equation}
Since sources use the same nested lattice codes, we can choose:
\begin{equation}
\SNR_{\mbox{\tiny{eff}}}=\min\{\SNR(\Vm_{k},\Am_{k}):k=1,2\}
\end{equation} where $\SNR(\Vm_{k},\Am_{k}) = M\SNR/\trace(\Vm_{k}\Am_{k}\Am_{k}^{\herm}\Vm_{k}^{\herm})$.

The decoding procedure is as follows.
We first consider the {\em aligned} received signals. Relay 1 observes
\begin{eqnarray}
\underline{\Ym}_{\mbox{\tiny{R}}_{1}} &=& \Fm_{11}\underline{\Xm}_{1}+\Fm_{12}\underline{\Xm}_{2} + \underline{\Zm}_{\mbox{\tiny{R}}_{1}}\nonumber\\
&\stackrel{(a)}{=}&\underbrace{\Fm_{11}\Vm_{1}}_{\triangleq \Hm_{\mbox{\tiny{R}}_{1}}}\left[
            \begin{array}{c}
              \underline{\tv}'_{1,1} \\
              \underline{\tv}'_{1,2} + \underline{\tv}'_{2,1} \\
              \vdots \\
              \underline{\tv}'_{1,M} + \underline{\tv}'_{2,M-1} \\
            \end{array}
          \right] + \underline{\Zm}_{\mbox{\tiny{R}}_{1}}\\
&=&\Hm_{\mbox{\tiny{R}}_{1}}\Cm_{\mbox{\tiny{R}}_{1}}\left[
                                \begin{array}{c}
                                  \underline{\Tm}_{1} \\
                                  \underline{\Tm}_{2} \\
                                \end{array}
                              \right]
                                  + \underline{\Zm}_{\mbox{\tiny{R}}_{1}} \label{eq:alignedsig1}
\end{eqnarray} where $(a)$ follows from the fact that the precoding vectors satisfy the alignment conditions in (\ref{cond:ALI}) and $\Cm_{\mbox{\tiny{R}}_{1}}=[
             \begin{array}{cc}
               \Am_{1} & \Cm_{12}\Am_{2} \\
             \end{array}]$ with
\begin{equation}
\Cm_{12}=\left[
                                                                \begin{array}{c}
                                                                  \mbox{0}^{1 \times M-1} \\
                                                                  \Id^{M-1 \times M-1} \\
                                                                \end{array}
                                                              \right].\label{eq:C12}
\end{equation}
Similarly, relay 2 observes the aligned signals:
\begin{eqnarray}
\underline{\Ym}_{\mbox{\tiny{R}}_{2}}
&=& \underbrace{\Fm_{21}\Vm_{1}}_{\triangleq \Hm_{\mbox{\tiny{R}}_{2}}}\left[
            \begin{array}{c}
              \underline{\tv}'_{1,1} + \underline{\tv}'_{2,1} \\
              \vdots \\
              \underline{\tv}'_{1,M-1} + \underline{\tv}'_{2,M-1} \\
               \underline{\tv}'_{1,M} \\
            \end{array}
          \right] + \underline{\Zm}_{\mbox{\tiny{R}}_{2}}\\
&=& \Hm_{\mbox{\tiny{R}}_{2}}\Cm_{\mbox{\tiny{R}}_{2}}\left[
                                \begin{array}{c}
                                  \underline{\Tm}_{1} \\
                                  \underline{\Tm}_{2} \\
                                \end{array}
                              \right]+\underline{\Zm}_{\mbox{\tiny{R}}_{2}}\label{eq:alignedsig2}
\end{eqnarray}where  $\Cm_{\mbox{\tiny{R}}_{2}}=[
             \begin{array}{cc}
               \Am_{1} & \Cm_{22}\Am_{2} \\
             \end{array}]$ with
\begin{equation}
\Cm_{22}=\left[
                                                                \begin{array}{c}
                                                                  \Id^{M-1 \times M-1} \\
                                                                  \mbox{0}^{1 \times M-1} \\
                                                                \end{array}
                                                              \right].\label{eq:C22}
\end{equation} The channel matrices in (\ref{eq:alignedsig1}) and (\ref{eq:alignedsig2}) follows the form considered in Section \ref{subsec:CoF}. Following the CoF framework in (\ref{cofrate}) and (\ref{cofeq}), if $R\leq R(\Hm_{\mbox{\tiny{R}}_{k}}\Cm_{\mbox{\tiny{R}}_{k}},\Bm_{\mbox{\tiny{R}}_{k}},\SNR_{\mbox{\tiny{eff}}})$, the relay $k$ can decode the $M$ linear combinations with full-rank coefficients matrix $\Bm_{\mbox{\tiny{R}}_{k}}$:
\begin{eqnarray*}
\underline{\Um}_{k} &=& [\Bm_{\mbox{\tiny{R}}_{k}}]_{q}[\Cm_{\mbox{\tiny{R}}_{k}}]_{q}\left[
                                          \begin{array}{c}
                                            \underline{\Wm}'_{1} \\
                                            \underline{\Wm}'_{2} \\
                                          \end{array}
                                        \right]\\
&=& [\Bm_{\mbox{\tiny{R}}_{k}}]_{q}\left[
     \begin{array}{cc}
       [\Am_{1}]_{q} & [\Cm_{k2}]_{q}[\Am_{2}]_{q} \\
     \end{array}
   \right]\left[
                                          \begin{array}{c}
                                            \underline{\Wm}'_{1} \\
                                            \underline{\Wm}'_{2} \\
                                          \end{array}
                                        \right]\\
&\stackrel{(a)}{=}&
   [\Bm_{\mbox{\tiny{R}}_{k}}]_{q}\left[
     \begin{array}{cc}
       \Id^{M \times M} & [\Cm_{k2}]_{q} \\
     \end{array}
   \right]\left[
                                          \begin{array}{c}
                                            \underline{\Wm}_{1} \\
                                            \underline{\Wm}_{2} \\
                                          \end{array}
                                        \right]
\end{eqnarray*} where $(a)$ is due to the precoding over $\FF_{q}$ in (\ref{eq:precoding}). Let $\hat{\underline{\Wm}}_{1} = [\Bm_{\mbox{\tiny{R}}_{1}}]_{q}^{-1}\underline{\Um}_{1}$ and $\hat{\underline{\Wm}}_{2}$ be the first $M-1$ rows of $[\Bm_{\mbox{\tiny{R}}_{2}}]_{q}^{-1}\underline{\Um}_{2}$. Then, we can define the deterministic noiseless finite field IC:
\begin{equation}
\left[
  \begin{array}{c}
    \hat{\underline{\Wm}_{1}} \\
    \hat{\underline{\Wm}_{2}} \\
  \end{array}
\right]=\Qm\left[
  \begin{array}{c}
    \underline{\Wm}_{1} \\
    \underline{\Wm}_{2} \\
  \end{array}
\right]\label{eq:fmodel}
\end{equation} where the so-called {\em system matrix} is obtained using $\Cm_{1}$ and $\Cm_{2}$ as
\begin{equation}
\Qm=\left[
      \begin{array}{cc}
        \Id^{M\times M} & \Qm_{12} \\
        \Qm_{21} & \Id^{M-1 \times M-1} \\
      \end{array}
    \right]\label{def:Q}
\end{equation} where
\begin{equation*}
\Qm_{12} = \left[
                                                                \begin{array}{c}
                                                                  \mbox{0}^{1 \times M-1} \\
                                                                  \Id^{M-1 \times M-1} \\
                                                                \end{array}
                                                              \right], \Qm_{21}= [
                                 \begin{array}{cc}
                                   \Id^{M-1 \times M-1} & \mbox{0}^{M-1 \times 1} \\
                                 \end{array}].
\end{equation*}

\subsection{Linear precoding over deterministic network}

Eq. (\ref{eq:fmodel}) defines the first hop of the noiseless finite field IC.
Next, we focus on the second hop.  The relay $k$ uses a precoded version of decoded linear  combinations
$\underline{\Wm}_{\mbox{\tiny{R}}_{k}}=\Mm_{k}\hat{\underline{\Wm}}_{k}$ as its messages.
Operating in a similar way as for the first hop,  the second hop noiseless finite field IC is given by
\begin{eqnarray}
\left[
                                             \begin{array}{c}
                                              \hat{\underline{\Wm}}_{\mbox{\tiny{R}}_{1}} \\
                                               \hat{\underline{\Wm}}_{\mbox{\tiny{R}}_{2}} \\
                                             \end{array}
                                           \right]&=&\Qm\left[
                                             \begin{array}{c}
                                               \underline{\Wm}_{\mbox{\tiny{R}}_{1}} \\
                                               \underline{\Wm}_{\mbox{\tiny{R}}_{2}} \\
                                             \end{array}
                                           \right]\label{eq:ff4}.
\end{eqnarray}
Concatenating (\ref{eq:fmodel}) and (\ref{eq:ff4}), the end-to-end finite field noiseless network is described by
\begin{equation}
\left[
                                             \begin{array}{c}
                                              \hat{\underline{\Wm}}_{\mbox{\tiny{R}}_{1}} \\
                                               \hat{\underline{\Wm}}_{\mbox{\tiny{R}}_{2}} \\
                                             \end{array}
                                           \right]=\Qm\left[
                                      \begin{array}{cc}
                                        \Mm_{1} & 0 \\
                                        0 & \Mm_{2} \\
                                      \end{array}
                                    \right]\Qm\left[
                                             \begin{array}{c}
                                               \underline{\Wm}_{1} \\
                                               \underline{\Wm}_{2} \\
                                             \end{array}
                                           \right].
\end{equation} Lemma \ref{lem:pre} shows that the linear combinations decoded at destination 1 are equal to its desired messages and
are equal to the messages with a change of sign (multiplication by $-1$ in the finite field)
at destination 2 (see Fig.~\ref{field}).
Notice that the system matrix is fixed and independent of the channel matrices, since it is determined only by the
alignment  conditions.

\begin{lemma}\label{lem:pre}
Choosing precoding matrices $\Mm_{1}$ and $\Mm_{2}$ as
\begin{eqnarray}
\Mm_{1}&=&(\Id^{M \times M}\oplus(-\Qm_{12}\Qm_{21}))^{-1}\\
\Mm_{2}&=& -(\Id^{M-1 \times M-1}\oplus(-\Qm_{21}\Qm_{12}))^{-1}
\end{eqnarray} the end-to-end system matrix becomes a diagonal matrix:
\begin{eqnarray*}
\Qm\left[
                                      \begin{array}{cc}
                                        \Mm_{1} & 0 \\
                                        0 & \Mm_{2} \\
                                      \end{array}
                                    \right]\Qm &=& \left[
                                      \begin{array}{cc}
                                        \Id^{M \times M} & 0 \\
                                        0 & -\Id^{M-1 \times M-1} \\
                                      \end{array}
                                    \right].
\end{eqnarray*}
\end{lemma}
\begin{IEEEproof}
See the long version of this paper \cite{Hong222}.
\end{IEEEproof}

Based on the above, we proved the following:
\begin{theorem}\label{thm} For the $2\times 2 \times 2$ MIMO IC defined in (\ref{model:1hop}) and (\ref{model:2hop}), Aligned PCoF can achieve the {\em symmetric} sum rate of $(2M-1)R$ with common message rate
\begin{equation*}
R=\min_{k=1,2}\{R(\Hm_{\mbox{\tiny{R}}_{k}}\Cm_{\mbox{\tiny{R}}_{k}},\Bm_{\mbox{\tiny{R}}_{k}},\SNR_{\mbox{\tiny{eff}}}),R(\Hm_{k}\Cm_{k},\Bm_{k},\SNR'_{\mbox{\tiny{eff}}})\}
\end{equation*} for any full rank integer matrices $\Am_{k},\Am_{\mbox{\tiny{R}}_{k}}$, $\Bm_{k},\Bm_{\mbox{\tiny{R}}_{k}}$, and any matrices $\Vm_{k},\Vm_{\mbox{\tiny{R}}_{k}}$ to satisfy the {\em alignment conditions} in (\ref{cond:ALI}), where
\begin{eqnarray*}
\Hm_{\mbox{\tiny{R}}_{k}} &=& \Fm_{k1}\Vm_{1}, \;\;\;\; \Hm_{k} = \Gm_{k1}\Vm_{\mbox{\tiny{R}}_{1}}\\
\SNR_{\mbox{\tiny{eff}}}&=&\min\{\SNR(\Vm_{k},\Am_{k}): k=1,2\}\\
\SNR'_{\mbox{\tiny{eff}}}&=&\min\{\SNR(\Vm_{\mbox{\tiny{R}}_{k}},\Am_{\mbox{\tiny{R}}_{k}}): k=1,2\},
\end{eqnarray*} and $\Cm_{\mbox{\tiny{R}}_{k}}=[\Am_{1} \mbox{     }\Cm_{k2}\Am_{2}]$ and $\Cm_{k}=[\Am_{\mbox{\tiny{R}}_{1}}\mbox{     }\Cm_{k2}\Am_{\mbox{\tiny{R}}_{2}}]$ with $\Cm_{12},\Cm_{22}$ in (\ref{eq:C12}) and (\ref{eq:C22}).\hfill \IEEEQED
\end{theorem}
Showing that $R$ grows as $\log \SNR$ yields:
\begin{corollary}  Aligned PCoF achieves the $2M-1$ DoF for the $2\times 2 \times 2$ MIMO IC when all nodes have $M$ multiple antennas.
\end{corollary}
\begin{IEEEproof}
See the long version of this paper \cite{Hong222}.
\end{IEEEproof}

\section{Optimization of symmetric sum rates}\label{sec:finite}

Suppose that precoding matrices $\Vm_{k},\Vm_{\mbox{\tiny{R}}_{k}}$ are determined. We need to optimize all integer matrices in Theorem \ref{thm} to maximize the sum rates. First, the power-penalty optimization problems take on the form:
\begin{eqnarray}
\argmin && \trace\left(\Vm\Am\Am^{\herm}\Vm^{\herm}\right)=\sum_{\ell=1}^{M}\|\Vm\av_{\ell}\|^2\nonumber\\
\mbox{subject to}&& \mbox{$\Am$ is full rank}\label{eq:opt1}
\end{eqnarray} where $\av_{\ell}$ denotes the $\ell$-th column of $\Am$. Also, the minimization problem of variance of effective noise consists of finding an integer matrix $\Bm$ solution of:
\begin{eqnarray}
\argmin && \max _{\ell}\{\bv_{\ell}^{\herm}(\SNR_{\mbox{\tiny{eff}}}^{-1}\Id+\Hm^{\herm}\Hm)^{-1}\bv_{\ell}\}\nonumber\\
\mbox{subject to}&& \mbox{$\Bm$ is full rank} \label{eq:opt2}
\end{eqnarray}where $\Hm$ denotes an aligned channel matrix and  $\bv_{\ell}^{\herm}$ is the $\ell$-th row of $\Bm$. By Cholesky decomposition, there exists a lower triangular matrix $\Lm$ such that
\begin{equation}
\bv_{\ell}^{\herm}(\SNR_{\mbox{\tiny{eff}}}^{-1}\Id+\Hm^{\herm}\Hm)^{-1}\bv_{\ell} = \|\Lm^{\herm}\bv_{\ell}\|^{2}.
\end{equation} We notice that problem (\ref{eq:opt1}) (or (\ref{eq:opt2})) is equivalent to finding a reduced basis for the lattice generated by $\Vm$ (or $\Lm^{\herm}$). In particular, the reduced basis takes on the form $\Vm\Um$ where $\Um$ is a unimodular matrix over $\ZZ[j]$. Hence, choosing $\Am=\Um$ yields the minimum power-penalty subject to the full rank condition in (\ref{eq:opt1}). In practice we used the (complex) LLL algorithm, with refinement of the LLL reduced basis approximation by Phost or Schnorr-Euchner lattice search (see the long version for details \cite{Hong222}).

\subsection{Numerical Results: Rayleigh fading}\label{subsec:NR}

\begin{figure}
\centerline{\includegraphics[width=9.5cm]{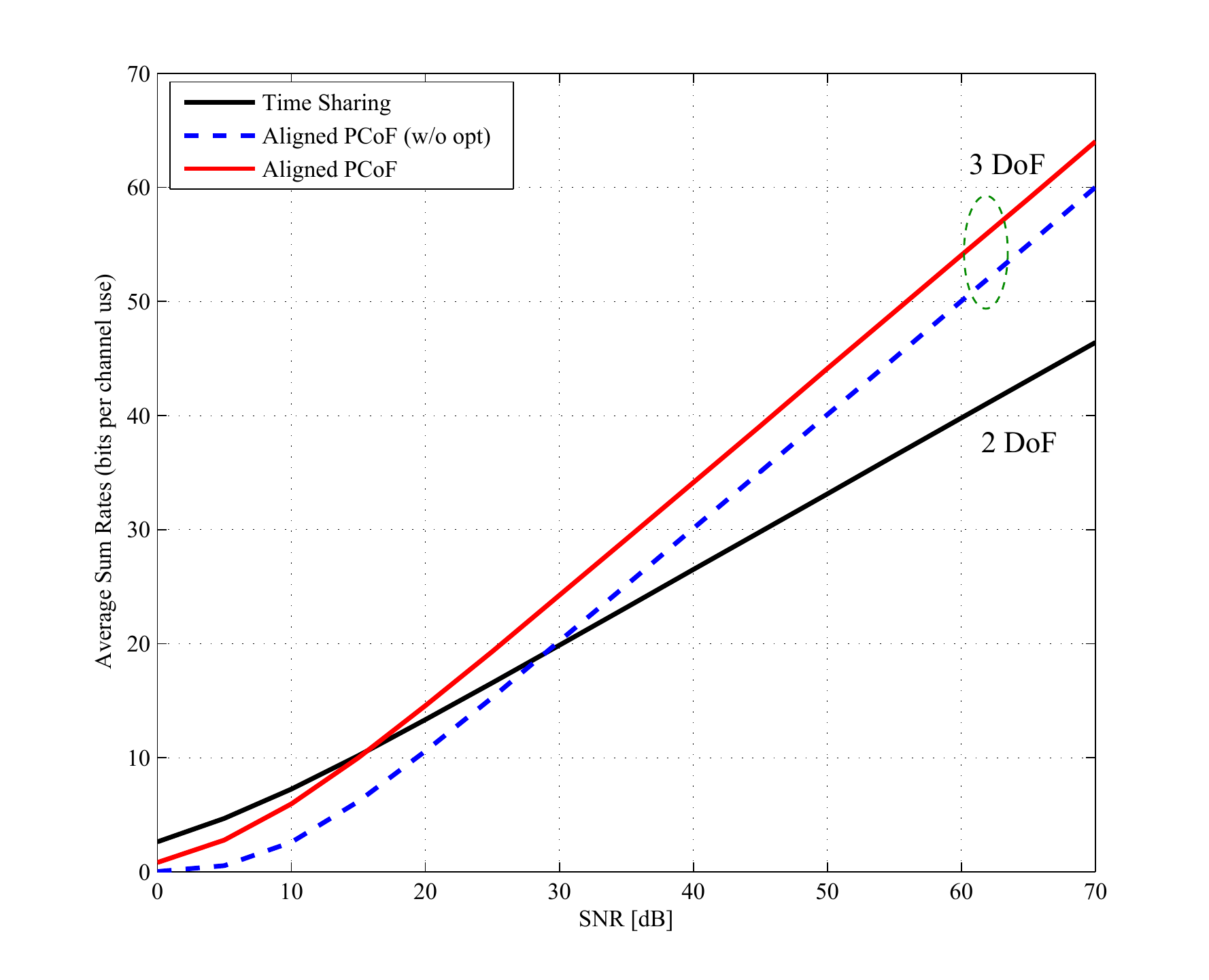}}
\caption{Performance comparison of Aligned PCoF and time-sharing with respect to ergodic symmetric sum rates.}
\label{simulation}
\end{figure}

We evaluate the performance of Aligned PCoF in terms of its average achievable sum rates. We computed the {\em ergodic} sum rates by Monte Carlo averaging with respect to the channel realizations with i.i.d. Rayleigh fading $\sim \Cc\Nc(0,1)$. For comparison, we considered the performance of {\em time-sharing} where IFR is used for each $M \times M$ MIMO IC. We used the IFR since it is known to almost achieve the performance of joint maximum likelihood receiver \cite{Zhan} and has a similar complexity with Aligned PCoF. In this case, an achievable symmetric sum rate is obtained as
\begin{equation}
R=\min\{R(\Fm_{kk},\Bm_{1},2\SNR), R(\Gm_{kk},\Bm_{2},2\SNR)\}
\end{equation}for any full-rank matrices $\Bm_{1}$ and $\Bm_{2}$. The integer matrices are optimized in the same manner of Aligned PCoF. Also, in this case, we used the $2M\SNR$ for power-constraint since each transmitter is active on every odd (or even) time slot. For Aligned PCoF, we need to find precoding matrices for satisfying the alignment condition in (\ref{cond:ALI}). For $M=2$, the conditions are given by
\begin{eqnarray}
\Fm_{11}\vv_{1,2} = \Fm_{12}\vv_{2,1} \mbox{ and } \Fm_{21}\vv_{1,1} = \Fm_{22}\vv_{2,1}.
\end{eqnarray} For the simulation, we used the following precoding matrices to satisfy the above conditions:
\begin{eqnarray*}
\Vm_{1} &=& \left[
              \begin{array}{cc}
                \Fm_{21}^{-1}\Fm_{12}\onev & \Fm_{11}^{-1}\Fm_{22}\onev \\
              \end{array}
            \right]
\mbox{ and } \vv_{2,1} =\onev.
\end{eqnarray*} Also, the same construction method is used for the second hop. Since source 1 (or relay 1) transmits one more stream than source 2 (or relay 2), the former always requires higher transmission power. In order to efficiently satisfy the average power-constraint, the role of sources 1 and 2 (equivalently, relays 1 and 2) is alternatively reversed in successive time slots. In Fig.~\ref{simulation}, we observe that Aligned PCoF can have the SNR gain about $5$ dB by optimizing the integer matrices for IFR and IFB, comparing with simply using identity matrices. Also, Aligned PCoF provides a higher sum rate than time-sharing if $\SNR \geq 15$ dB,
and its gain over time-sharing increases with $\SNR$, showing that in this case the DoF result matters also at finite SNR.

\end{document}